\documentclass[aps,prl,reprint]{revtex4-1}

\usepackage[english]{babel}

\usepackage{amsmath,amssymb}

\usepackage{textcomp}

\usepackage{graphics}

\usepackage[T1]{fontenc}

\begin{document}

\title{A continuum theory of phase separation kinetics for active Brownian particles}

\author{Joakim Stenhammar}
\email{j.stenhammar@ed.ac.uk}
\affiliation{SUPA, School of Physics and Astronomy, University of Edinburgh, JCMB Kings Buildings, Edinburgh EH9 3JZ, United Kingdom}

\author{Adriano Tiribocchi}
\affiliation{SUPA, School of Physics and Astronomy, University of Edinburgh, JCMB Kings Buildings, Edinburgh EH9 3JZ, United Kingdom}

\author{Rosalind J. Allen}
\affiliation{SUPA, School of Physics and Astronomy, University of Edinburgh, JCMB Kings Buildings, Edinburgh EH9 3JZ, United Kingdom}

\author{Davide Marenduzzo}
\affiliation{SUPA, School of Physics and Astronomy, University of Edinburgh, JCMB Kings Buildings, Edinburgh EH9 3JZ, United Kingdom}

\author{Michael E. Cates}
\affiliation{SUPA, School of Physics and Astronomy, University of Edinburgh, JCMB Kings Buildings, Edinburgh EH9 3JZ, United Kingdom}

\date{\today}


\begin{abstract} 
Active Brownian particles (ABPs), when subject to purely repulsive interactions, are known to undergo activity-induced phase separation broadly resembling an equilibrium (attraction-induced) gas-liquid coexistence. Here we present an accurate continuum theory for the dynamics of phase-separating ABPs, derived by direct coarse-graining, capturing leading-order density gradient terms alongside an effective bulk free energy. Such gradient terms do not obey detailed balance; yet we find coarsening dynamics closely resembling that of equilibrium phase separation. Our continuum theory is numerically compared to large-scale direct simulations of ABPs and accurately accounts for domain growth kinetics, domain topologies and coexistence densities. 
\end{abstract}

\maketitle

Active matter -- materials whose constituents convert energy from an internal or external fuel depot into work -- has gathered significant attention over the last decade \cite{Ramaswamy-2010,Romanczuk-2012,Marchetti-2013}. One important paradigm for active matter is a fluid of self-propelled particles. These can be natural,  e.g., a bacterial or algal suspension \cite{Cates-2012,Marchetti-2013}, or man-made, such as colloidal particles rendered motile through chemical reactions \cite{Howse-2007,Ebbens-2010,Volpe-2011,Thutupalli-2011,Palacci-2013}. Such colloids swim at roughly constant speed, with a swimming direction that relaxes continuously by rotational diffusion; this defines `active Brownian particles' (ABPs).

\begin{figure}[h!] 
\begin{center}
\resizebox{!}{115mm}{\includegraphics{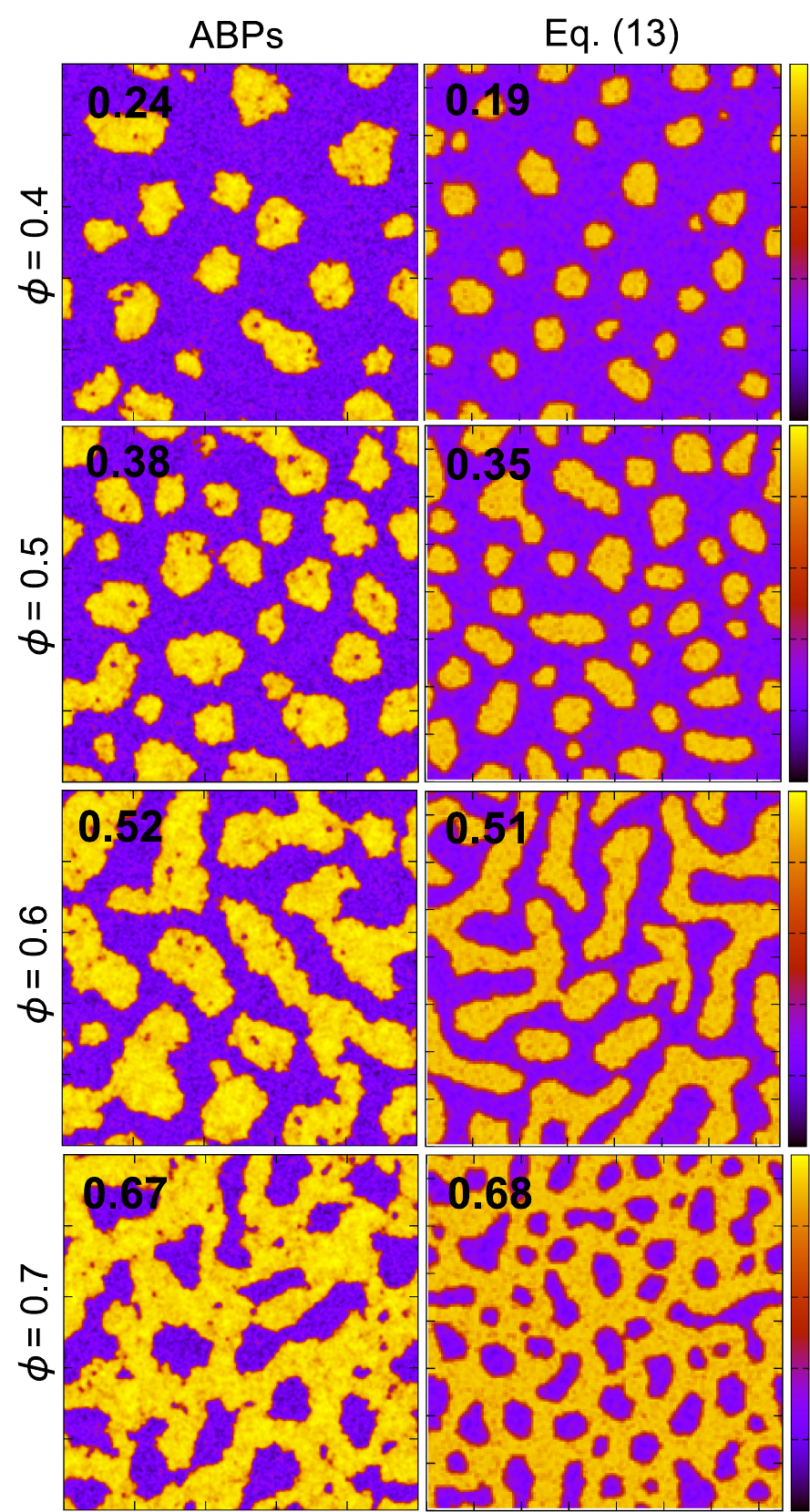}}
\caption{ABP (left) and continuum (right) simulations with particle area fractions $\phi$ as indicated. Snapshots are taken at equal time $t=500\tau_{\mathrm{r}}$ and shown at the same spatial scale. Inset numbers indicate the area fraction of the dense phase, measured through numerical integration of $P(\phi)$ (Fig.~\ref{phi_prob}). The color bars run from $\phi=0$ (black) to $\phi=1$ (yellow).}
\label{snapshots}
\end{center}
\end{figure}

Active matter represents an inherently far-from-equilibrium system. This causes a range of nontrivial behaviors, such as giant density fluctuations \cite{Narayan-2007,Deseigne-2010,Zhang-2010,Wensink-2012}, rectification of motion \cite{Galajda-2007,Angelani-2009,Lambert-2010,Potosky-2013}, and unexpected phase separations \cite{Schweitzer-1994,Cates-2010,Theurkauff-2012,McCandlish-2012,Farrell-2012,Redner-2013-PRE,Buttinoni-2013}. Of particular relevance to us is the prediction that a suspension of motile particles with a density-dependent swim speed $v(\rho)$, which decreases with increasing density due to crowding, can phase separate even without attractive interactions or orientational order~\cite{Tailleur-2008,Cates-2013}. This effect relies on the fact that both run-and-tumble bacteria~\cite{Tailleur-2008,Schnitzer-1993}, and ABPs~\cite{Cates-2013}, accumulate in regions where they move slowly. A positive feedback, whereby a local density increase leads to a local slowdown, causes further accumulation. This motility-induced phase separation has been confirmed in simulations (resembling those in the left panel of Fig. \ref{snapshots}) \cite{Fily-2012,Redner-2013} and experiments \cite{Buttinoni-2013} and shares many features with the equilibrium gas-liquid coexistence of passive attractive particles, even though this feedback mechanism is completely absent in systems obeying detailed balance. In ABP systems, a decreasing $v(\rho)$ effectively arises from an increased collision frequency in dense regions. As shown in Ref.~\cite{Redner-2013} (see further Fig. S1 of \cite{SI}), phase separation is described by a gas-liquid-like phase diagram where the traditional role of the inverse temperature is played by the P\'eclet number, $\mathrm{Pe}=3v_0\tau_{\mathrm{r}}/\sigma$, where $v_0 = v(0)$ is the propulsion speed of an isolated ABP, $\sigma$ its diameter, and $\tau_{\mathrm{r}}$ its orientational relaxation time.

While previous studies~\cite{Fily-2012,Redner-2013} have focused on steady-state properties, in this Letter we investigate how far the correspondence between thermal and athermal phase separation extends to \emph{dynamics} by developing a continuum description for the structural evolution. We explicitly coarse-grain the microscopic dynamics into an evolution equation for the density field $\rho(\mathbf{r})$. This contains an effective diffusivity and chemical potential that stem from a functional form of $v(\rho)$ suggested by kinetic arguments, and validated by our direct ABP simulations. Solving our continuum equation numerically we find domain growth dynamics and morphologies in very good accord with large-scale direct ABP simulations, while the efficiency of the continuum approach allows a two-decade extension to the simulated time window. Although the effective chemical potential violates detailed balance at square-gradient level, in practice we find the effects of this violation to be rather limited; our striking conclusion is that even the \emph{dynamics} of activity-induced phase separation in ABPs (a manifestly far-from-equilibrium effect) is quantitatively captured by a continuum model that only weakly transgresses the boundaries of equilibrium statistical thermodynamics. 

\emph{Derivation of the dynamical equation.} As derived in Ref~\cite{Tailleur-2008} and generalized in~\cite{Cates-2013}, the coarse-grained density field $\rho$ for ABPs with density-dependent swim speed obeys:
\begin{equation}\label{rho_t}
\partial_t \rho = -\nabla\cdot \left\{ -D(\rho) \rho \nabla \mu + \sqrt{2 D(\rho) \rho} {\boldsymbol \Lambda} \right\},
\end{equation}
Here $D(\rho)$ is an effective one-body diffusivity, $\mu$ an effective chemical potential, and ${\boldsymbol\Lambda}$ a noise vector whose Cartesian components $\Lambda_i$ obey $\langle \Lambda_i(\mathbf{r},t) \Lambda_j(\mathbf{r}',t') \rangle = \delta_{ij}\delta(\mathbf{r}-\mathbf{r}') \delta(t-t')$. The multiplicative noise is It\={o}-type as derived in Ref~\cite{Tailleur-2008}, which furthermore showed that when $v(\rho)$ is a strictly local function, 
\begin{equation}\label{mu_0}
\mu(\rho) = \mu_0(\rho) \equiv \ln\rho+\ln v(\rho).
\end{equation} 
As shown previously \cite{Tailleur-2008}, the local term $\mu_0$ can be written as the functional derivative of an effective bulk free energy $\mathcal{F}_0=\int f_0(\rho)\mathrm{d}\mathbf{r}$, where 
\begin{equation}\label{f_0}
f_0 = \rho (\ln \rho -1) + \int_{0}^{\rho} \ln \left[ v(u)\right] \mathrm{d}u.
\end{equation}
The first term resembles the standard ideal entropy of a passive fluid and the second its excess free-energy density. The latter is of similar form to an enthalpic attraction (despite its completely different physical origins) and causes bulk phase separation for steeply enough decreasing $v(\rho)$. 

While the above treatment predicts the \emph{existence} of phase separation, an extension to non-trivial order in density gradients is necessary to stabilize domain walls and thereby enable the study of phase separation \emph{dynamics}. We thus assume that $v$ is not strictly local, but that ABPs sample $\rho$ on a length scale significantly greater than the interparticle spacing. Therefore we set $v=v(\hat\rho)$ with $\hat\rho(\mathbf{r})= \rho+\gamma^2\nabla^2\rho$ and $\gamma(\rho)$ a smoothing length: this represents the leading order nonlocal correction to $\rho$ allowed by rotational invariance. Replacing $v(\rho)$ by $v(\hat\rho)$ and assuming that $\gamma$ is proportional to the persistence length of ABP trajectories (\emph{i.e.}, $\gamma(\rho)=\gamma_0\tau_{\mathrm{r}}v(\rho)$, where $\gamma_0$ is of order unity) we find
\begin{align}
\mu &= \mu_0 - \kappa(\rho)\nabla^2\rho + \mathcal{O}(\nabla^{4}\rho), \label{mu_full} \\
\kappa(\rho) &\equiv -\gamma_{0}^{2} \tau_{\mathrm{r}}^{2} v(\rho) \frac{\mathrm{d} v(\rho)}{\mathrm{d} \rho}.\label{kappa}
\end{align}
Thus our microscopic arguments point to a specific form of the square gradient term in \eqref{mu_full}, different from the phenomenological assumption of constant $\kappa$ (\emph{e.g.},~\cite{Cates-2010}). 

The gradient structure of our effective chemical potential at first sight resembles that for an interfacial free energy density in passive systems, $f_{\mathrm{int}}=(\kappa/2)(\nabla\rho)^{2}$. However, on functional differentiation to obtain $\mu$, the latter would lead not to \eqref{mu_full} but to
\begin{equation}\label{DB_intact}
\mu = \mu_0 - \kappa(\rho)\nabla^2\rho - \frac{\mathrm{d}\kappa(\rho)}{\mathrm{d}\rho}\frac{(\nabla\rho)^2}{2}.
\end{equation}
Our microscopic analysis hence demonstrates that, even when the local contribution $\mu_0$ maps onto an effective free energy~\cite{Tailleur-2008}, square-gradient terms in active systems generally do not.

Finally, to emulate the physics of excluded volume interactions between ABPs at the continuum level, we add a stabilizing contribution to the effective free energy density, $f_0\to f_0+f_{\mathrm{rep}}$, that increases sharply above a threshold density $\rho_{\mathrm{t}}$. We choose
\begin{equation}\label{f_rep}
f_{\mathrm{rep}} = k_{\mathrm{rep}} \Theta(\rho - \rho_{\mathrm{t}})(\rho-\rho_{\mathrm{t}})^{4}
\end{equation} 
with $\Theta(x)$ the step function and $k_{\mathrm{rep}}$ a phenomenological parameter. While the form of $f_{\mathrm{rep}}$ somewhat affects the phase boundaries and the coexistence densities (Fig. S2 in \cite{SI}), we have checked that the structure and dynamics of the phase-separating system, which are our focus here, are not sensitive to this choice. 

\emph{Microscopic estimate of density-dependent speed.} On time scales larger than the orientational relaxation time $\tau_{\mathrm{r}}$, an isolated ABP undergoes a persistent random walk of effective step length $\ell_0\simeq v_0 \tau_{\mathrm{r}}$. At non-zero density, collisions slow particles down, giving an effective propulsion speed $v(\rho)<v_0$. To estimate this, assume that each particle has velocity $v_0$ between collisions, but is effectively stalled for the duration $\tau_{\mathrm{c}}$ of each collision event. Since collisions do not change $\tau_{\mathrm{r}}$ itself (a picture that may change significantly if hydrodynamic interactions are added \citep{Fielding-2012}), the length $\ell(\rho)$ traveled during $\tau_{\mathrm{r}}$ obeys $\ell(\rho)=v_0 (\tau_{\mathrm{r}}-n_{\mathrm{c}}\tau_{\mathrm{c}})$ where $n_{\mathrm{c}}$ is the average number of collisions during $\tau_{\mathrm{r}}$. This leads to an effective propulsion speed $v(\rho)=\ell(\rho)/\tau_{\mathrm{r}}=v_0(1-n_{\mathrm{c}} \tau_{\mathrm{c}}/\tau_{\mathrm{r}})$. We now write $n_{\mathrm{c}}=\tau_{\mathrm{r}}/(\tau_{\mathrm{MF}}+\tau_{\mathrm{c}})$, where $\tau_{\mathrm{MF}}$ is the mean free time between collisions, so $(\tau_{\mathrm{MF}}+\tau_{\mathrm{c}})$ is the time between the starts of two adjacent collision events. Furthermore assuming that $\tau_{\mathrm{MF}}\gg\tau_{\mathrm{c}}$, which is valid at low densities, we get
\begin{equation}\label{v_eff_final}
v(\rho) \approx v_0 \left(1 - \frac{\tau_{\mathrm{c}}}{\tau_{\mathrm{MF}}}\right) = v_0 (1 - v_0 \sigma_{\mathrm{s}} \tau_{\mathrm{c}} \rho  ),
\end{equation}
where we have used a standard expression $\tau_{\mathrm{MF}}=(v_0\rho\sigma_{\mathrm{s}})^{-1}$ from kinetic theory, with $\sigma_{\mathrm{s}}$ a scattering cross section. Finally, within the density range where this collision-hampered random walk picture remains valid, the effective diffusivity $D(\rho)$ obeys in two dimensions~\cite{Cates-2013}
\begin{equation}\label{D_eff}
D(\rho) = \frac{v^{2}(\rho)\tau_{\mathrm{r}}}{2} = D_0(1-v_0 \sigma_{\mathrm{s}} \tau_{\mathrm{c}}\rho)^{2},
\end{equation}
where $D_0\equiv D(0)=v_0^{2}\tau_\mathrm{r}/2$. 

\begin{figure}[h] 
\begin{center}
\resizebox{!}{50mm}{\includegraphics{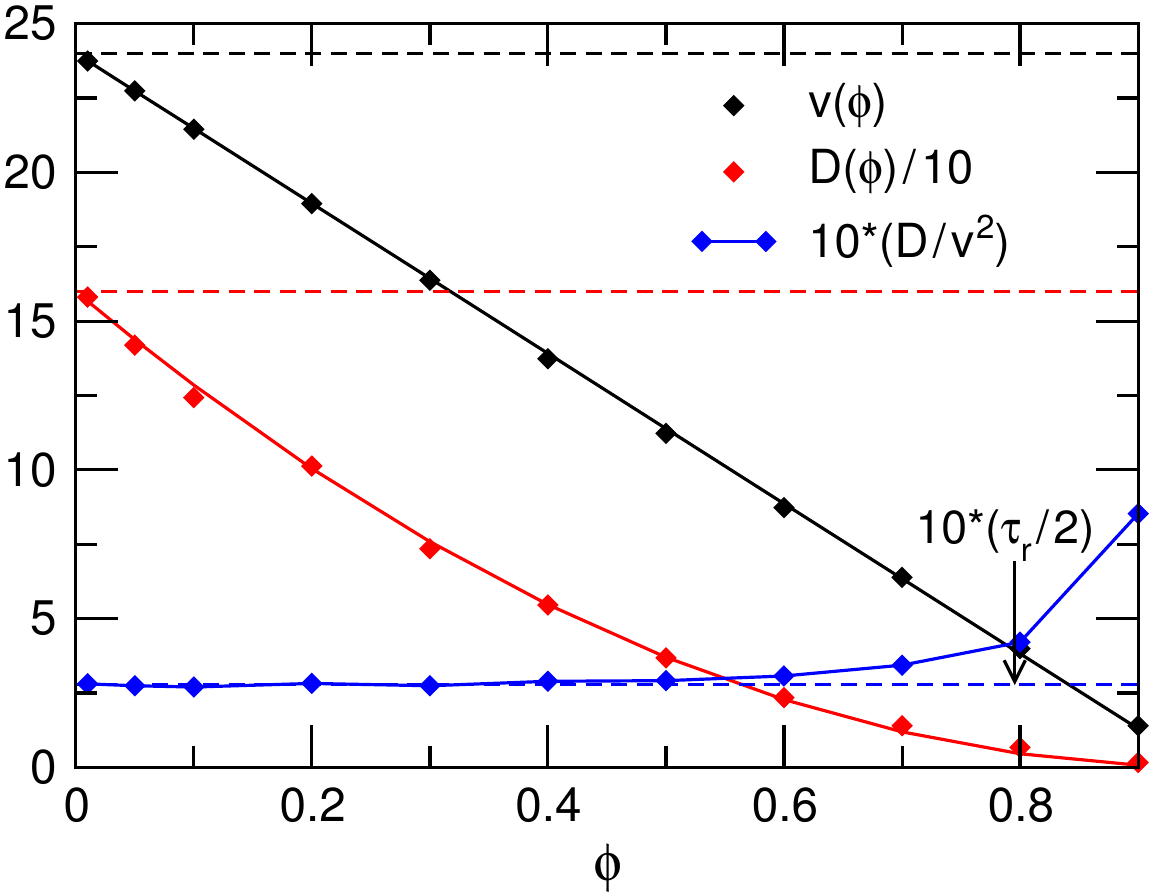}}
\caption{Density-dependent swim speed $v(\phi)$ (black symbols) and diffusivity $D(\phi)$ (red symbols), and the ratio $D/v^{2}$ (blue curve and symbols) obtained from an ABP simulation at Pe = 40. The black and red curves show the best fits to the functions $v_0(1-a\phi)$ and $D_0(1-b\phi)^{2}$, respectively, with the optimized values $a=1.05$ and $b=1.04$. Dashed lines show the predicted zero-density values. Plotted quantities are in Lennard-Jones units $\sigma$ and $\tau_{\mathrm{LJ}}$, as defined in \cite{SI}.}
\label{diff_rho}
\end{center}
\end{figure}

Figure \ref{diff_rho} shows $D(\phi)$ and $v(\phi)$ as functions of the particle area fraction $\phi\in [0.01,0.9]$ obtained from ABP simulations in the one-phase region (Pe = 40, see further Fig. S1 in \cite{SI}). Both measured quantities follow very closely the predictions of \eqref{v_eff_final} and \eqref{D_eff}. Furthermore, the ratio $D/v^{2}$ coincides with the predicted value of $\tau_{\mathrm{r}}/2$ for area fractions as high as $\phi\approx 0.7$. Fitting the data for $v(\phi)$ allows us to estimate the ratio $\tau_{\mathrm{MF}}/\tau_{\mathrm{c}}$ and leads to $\tau_{\mathrm{MF}}/\tau_{\mathrm{c}}\approx \phi^{-1}$; thus, the assumption $\tau_{\mathrm{MF}}\gg \tau_{\mathrm{c}}$ should fail for $\phi\gtrsim 0.5$, which is indeed where the ratio between $D$ and $v^{2}$ starts to deviate from its zero-density value. We also note that the optimized fitting parameters $a=1.05$ and $b=1.04$ in $v(\phi)=v_0(1-a\phi)$ and $D(\phi)=D_0(1-b\phi)^{2}$ are nearly identical, as expected from the kinetic reasoning above, and as and as was recently predicted theoretically \citep{Bialke-2013} and observed for a similar ABP model \citep{Fily-2012}. We finally note that the collision time $\tau_{\mathrm{c}}$ is in principle an increasing function of Pe, leading to a Pe-dependence in the fitting parameters $a$ and $b$. However, as shown in Fig. S3 of \cite{SI}, this dependence is weak throughout the range $40\leq \mathrm{Pe}\leq 100$ considered here. We thus assume that $v(\phi)$ and $D(\phi)$ obtained from the (homogeneous) $\mathrm{Pe}=40$ system are transferable to the $\mathrm{Pe}=100$ case where our kinetic studies will be conducted. 

\emph{Mapping between continuum and ABP models.} To allow direct comparison with particle-based ABP simulations we first rewrite the continuum model in terms of a dimensionless variable $\tilde{\phi}=\rho/\rho_0$ where $\rho_0=(v_0\sigma_{\mathrm{s}}\tau_{\mathrm{c}})^{-1}$ so that $v(\tilde{\phi})=v_0(1-\tilde{\phi})$. Matching to the ABP results of Fig.~\ref{diff_rho} shows $\tilde\phi \simeq 1.05\phi$ where $\phi$ is the area fraction of ABPs. For our purposes it is enough to ignore this small difference and use $\phi$ to denote both quantities. We will furthermore use $\lambda\equiv D_0/v_0$ and $\tau_{\mathrm{r}}/2=D_0/v_0^{2}$ as our units of length and time, respectively, for which it can be shown that $v_0 = D_0 = 1$. In these units, Eqs.~\eqref{mu_0}--\eqref{f_0}, \eqref{f_rep} furthermore become
\begin{align}\label{mu_0_final}
\mu_0 &= \ln \left[\phi(1-\phi) \right] 
\\
\mu_{\mathrm{rep}} &= 4k_{\mathrm{rep}}\Theta(\phi - \phi_{\mathrm{t}})(\phi - \phi_{\mathrm{t}})^{3}
\\
\label{mu_nabla_final}
\mu &= \mu_0 +\mu_{\mathrm{rep}}-\kappa_0(1-\phi)\nabla^{2}\phi,
\end{align}
where $\phi_{\mathrm{t}}=\rho_{\mathrm{t}} / \rho_0$ and $\kappa_0=(v_0\gamma_0\tau_{\mathrm{r}})^{2}=4\gamma_0^{2}$; since the order-unity factor $\gamma_0$ is unknown, we treat $\kappa_0$ as a density-independent free parameter (see \cite{SI} for parameter values). Finally, Eq.~\eqref{rho_t} becomes
\begin{equation}\label{rho_t_rescaled}
\partial_t \phi = \nabla\cdot \left\{\phi (1-\phi)^{2} \nabla \mu- \sqrt{2 \phi (1-\phi)^{2} N_0^{-1}} {\boldsymbol \Lambda} \right\}
\end{equation}
which we solve numerically using standard methods. The results are compared to large-scale ($N\approx 5\times 10^{5}$) Brownian dynamics simulations of repulsive ABPs using the LAMMPS package \citep{Plimpton-1995} (for model details, see \cite{SI}). The comparison is achieved by constructing a density field from the ABP simulations through numerical coarse-graining. With our choice of units, Eq. \eqref{rho_t} fixes directly the noise term in \eqref{rho_t_rescaled}, with $N_0$ being the number of ABPs in a cell of side $\lambda$ at nominal area fraction unity, a number which is readily accessible from the known parameters of the ABP simulations. Thus, since our choice of units sets an absolute scale of length and time matched across both simulations (and, as noted above, the results are insensitive to $k_{\mathrm{rep}}$), $\kappa_0$ remains the sole fit parameter in our comparison. 

\emph{Numerical results.} In Fig.~\ref{snapshots}, intermediate-time snapshots of the phase-separating system at four different particle area fractions are shown, obtained through solving the continuum model and by explicit simulations of ABPs. Clearly, the continuum model describes the domain structure very well, particularly in the middle of the density range; since the binodal is imperfectly reproduced, deviation at the extremes is expected. Qualitatively, the observed domain topologies resemble an equilibrium spinodal decomposition \cite{Velasco-1996,Novik-2000}, starting from isolated domains of dense phase at low $\phi$, \emph{via} a near-bicontinuous structure at $\phi=0.6$, to isolated droplets of dilute phase in a dense matrix at still larger $\phi$. Furthermore, the areas occupied by the two phases agree well with the ABP simulations. The quantitative agreement reported here is easily destroyed by deviating from the parameter mappings delineated above, for instance by setting $D(\phi)$ to a constant. We finally note that the noisy local dynamics observed in ABP simulations is qualitatively captured by the continuum model using the noise strength determined by our mapping (see movies in \cite{SI}).

\begin{figure}[h] 
\begin{center}
\resizebox{!}{50mm}{\includegraphics{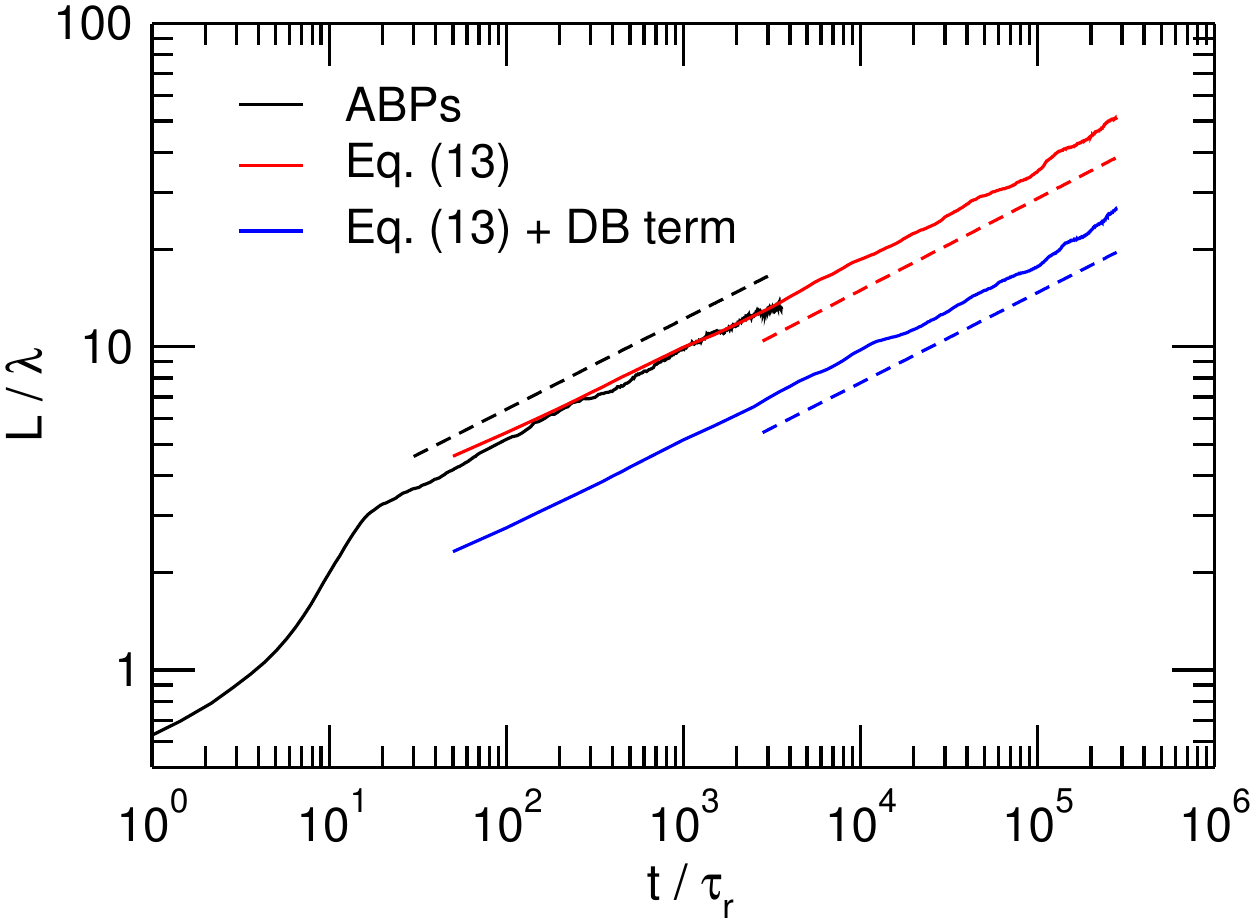}}
\caption{Time-dependent coarsening length $L(t)$ obtained from the inverse first moment of the structure factor at $\phi=0.5$. The dashed lines indicate the fitted exponents $\alpha=0.27(9)$ (ABPs), $\alpha=0.28(7)$ (continuum model), and $\alpha=0.27(9)$ (continuum model with detailed balance (DB) term). The latter curve has been vertically shifted for clarity.}
\label{L_t}
\end{center}
\end{figure}

Turning to kinetics, Fig.~\ref{L_t} shows the time-evolution of the characteristic length-scale $L(t)$ (as defined in \cite{SI}) for ABPs and for the continuum model. In equilibrium systems $L(t)$ usually exhibits power-law coarsening, $L(t)\sim t^{\alpha}$, where the growth exponent $\alpha$ depends on the kinetic universality class. For phase separations with diffusive transport of the order parameter and where hydrodynamic interactions can be neglected, one expects $\alpha=1/3$~\cite{Chaikin,Bray-2002}. Interestingly, our continuum model instead exhibits an exponent $\alpha\simeq 0.28$, close to the $0.255$ previously reported in Ref.~\cite{Redner-2013} (and later corrected to 0.272 \cite{Redner-email}). As well as giving the same scaling exponent, the ABP and continuum curves connect almost perfectly onto each other, demonstrating the quantitative accuracy of our continuum description. We finally note that the continuum theory enables us to extend the simulated time window by two decades beyond that possible for direct ABP simulations.

A more detailed analysis would be needed to understand the exact nature of this sub-diffusive domain growth, and indeed to confirm whether it represents a true asymptotic behavior (as observed for the Cahn-Hilliard equation with a density-dependent mobility similar to the one used here \cite{Bray-1995}) or a transient crossover (due for example to the high noise level ~\cite{Bray-2002}). Notably, however, detailed balance violations do not seem to be responsible for the exponent anomaly: a repeat run with detailed balance restored as per Eq.~\eqref{DB_intact} shows the same exponent within the numerical accuracy (Fig.~\ref{L_t}). 

\begin{figure}[h] 
\begin{center}
\resizebox{!}{50mm}{\includegraphics{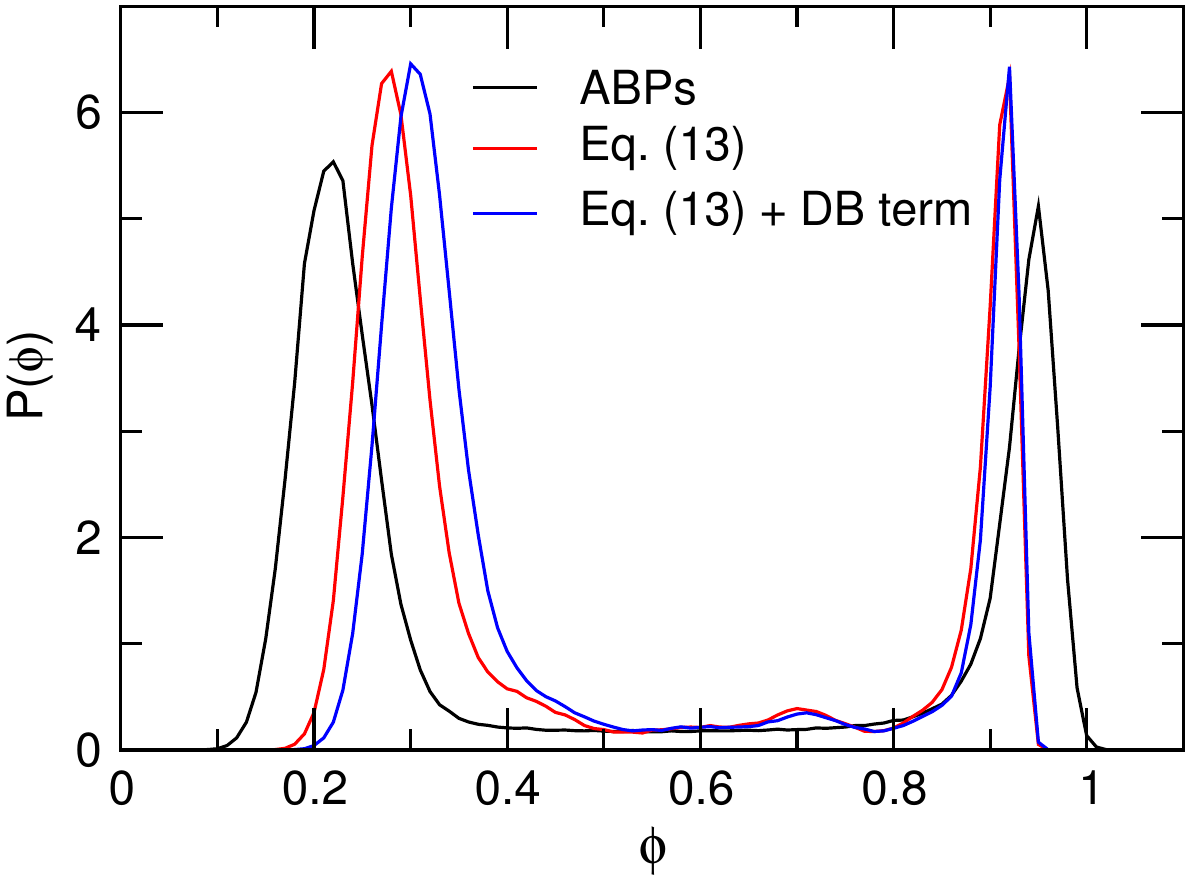}}
\caption{Probability distribution $P(\phi)$ of the local area fraction $\phi$ obtained from ABP simulations (black curve), from the continuum model as written (red curve) and with detailed balance restored (blue curve). $P(\phi)$ was sampled over quadratic coarse-graining areas of side length $0.8\lambda$ at $\phi=0.5$ and averaged over the time window $500\tau_{\mathrm{r}}\leq t\leq 3500\tau_{\mathrm{r}}$.}
\label{phi_prob}
\end{center}
\end{figure}

To further quantify the structure, we plot in Fig.~\ref{phi_prob} the probability distribution $P(\phi)$ mid-run during coarsening. Given the previously noted discrepancy between the coexistence densities, the agreement in shape between the $P(\phi)$ curves is good. 
Fig.~\ref{phi_prob} also shows that detailed balance violations at square-gradient level slightly shift the coexistence densities, marginally improving agreement with the ABP curve. This shift in binodals is not a numerical artefact, and its underlying physics will be pursued further in a separate publication; in the context of phase separation kinetics, however, the effects of detailed balance violations remain small. These findings are surprising: detailed balance violations enter our model only via gradient terms, which for systems with detailed balance have no effect on bulk phase equilibria yet are crucial in determining interfacial structure and hence coarsening behavior. Here we find almost the opposite. 

\emph{Conclusions and outlook.} By explicit coarse-graining we have derived a stochastic continuum model for active Brownian particles with repulsive collisional interactions. We have shown that this model quantitatively describes their phase separation dynamics, with essentially a single fit parameter. Although at square gradient level the model no longer respects detailed balance, in practice detailed balance violations have little influence on the observed microstructures and coarsening dynamics. It is surprising that the kinetics of a phase separation induced by an inherently non-equilibrium process is quantitatively so similar to that of a purely thermodynamic system -- albeit one with a carefully chosen free energy, mobility, and noise level. This encourages further work on the fundamental connections between equilibrium and non-equilibrium thermodynamics in the context of active matter. This seems essential if the rich phenomenology seen experimentally in both ABPs~\cite{Howse-2007,Volpe-2011,Palacci-2013,Theurkauff-2012,Buttinoni-2013} and bacterial systems~\cite{Galajda-2007,Angelani-2009,Lambert-2010,Zhang-2010,Cates-2010} is to be understood.

\begin{acknowledgments}
Helpful discussions with Alan Bray, Aidan Brown, Tom Lion, Julien Tailleur, and Raphael Wittkowski are gratefully acknowledged, and JS would like to thank Fred Farrell for assistance with the numerical coarse-graining. We thank EPSRC EP/J007404 for funding. JS is supported by the Swedish Research Council (350-2012-274), RJA by a Royal Society University Research Fellowship, and MEC by a Royal Society Research Professorship. 
\end{acknowledgments}

\bibliography{bibliography}

\begin{thebibliography}{39}%
\makeatletter
\providecommand \@ifxundefined [1]{%
 \@ifx{#1\undefined}
}%
\providecommand \@ifnum [1]{%
 \ifnum #1\expandafter \@firstoftwo
 \else \expandafter \@secondoftwo
 \fi
}%
\providecommand \@ifx [1]{%
 \ifx #1\expandafter \@firstoftwo
 \else \expandafter \@secondoftwo
 \fi
}%
\providecommand \natexlab [1]{#1}%
\providecommand \enquote  [1]{``#1''}%
\providecommand \bibnamefont  [1]{#1}%
\providecommand \bibfnamefont [1]{#1}%
\providecommand \citenamefont [1]{#1}%
\providecommand \href@noop [0]{\@secondoftwo}%
\providecommand \href [0]{\begingroup \@sanitize@url \@href}%
\providecommand \@href[1]{\@@startlink{#1}\@@href}%
\providecommand \@@href[1]{\endgroup#1\@@endlink}%
\providecommand \@sanitize@url [0]{\catcode `\\12\catcode `\$12\catcode
  `\&12\catcode `\#12\catcode `\^12\catcode `\_12\catcode `\%12\relax}%
\providecommand \@@startlink[1]{}%
\providecommand \@@endlink[0]{}%
\providecommand \url  [0]{\begingroup\@sanitize@url \@url }%
\providecommand \@url [1]{\endgroup\@href {#1}{\urlprefix }}%
\providecommand \urlprefix  [0]{URL }%
\providecommand \Eprint [0]{\href }%
\providecommand \doibase [0]{http://dx.doi.org/}%
\providecommand \selectlanguage [0]{\@gobble}%
\providecommand \bibinfo  [0]{\@secondoftwo}%
\providecommand \bibfield  [0]{\@secondoftwo}%
\providecommand \translation [1]{[#1]}%
\providecommand \BibitemOpen [0]{}%
\providecommand \bibitemStop [0]{}%
\providecommand \bibitemNoStop [0]{.\EOS\space}%
\providecommand \EOS [0]{\spacefactor3000\relax}%
\providecommand \BibitemShut  [1]{\csname bibitem#1\endcsname}%
\let\auto@bib@innerbib\@empty
\bibitem [{\citenamefont {Ramaswamy}(2010)}]{Ramaswamy-2010}%
  \BibitemOpen
  \bibfield  {author} {\bibinfo {author} {\bibfnamefont {S.}~\bibnamefont
  {Ramaswamy}},\ }\href@noop {} {\bibfield  {journal} {\bibinfo  {journal}
  {Annu. Rev. Condens. Matter Phys.}\ }\textbf {\bibinfo {volume} {1}},\
  \bibinfo {pages} {323} (\bibinfo {year} {2010})}\BibitemShut {NoStop}%
\bibitem [{\citenamefont {Romanczuk}\ \emph {et~al.}(2012)\citenamefont
  {Romanczuk}, \citenamefont {B\"ar}, \citenamefont {Ebeling}, \citenamefont
  {Lindner},\ and\ \citenamefont {Schimansky-Geier}}]{Romanczuk-2012}%
  \BibitemOpen
  \bibfield  {author} {\bibinfo {author} {\bibfnamefont {P.}~\bibnamefont
  {Romanczuk}}, \bibinfo {author} {\bibfnamefont {M.}~\bibnamefont {B\"ar}},
  \bibinfo {author} {\bibfnamefont {W.}~\bibnamefont {Ebeling}}, \bibinfo
  {author} {\bibfnamefont {B.}~\bibnamefont {Lindner}}, \ and\ \bibinfo
  {author} {\bibfnamefont {L.}~\bibnamefont {Schimansky-Geier}},\ }\href@noop
  {} {\bibfield  {journal} {\bibinfo  {journal} {Eur. Phys. J. Special Topics}\
  }\textbf {\bibinfo {volume} {202}},\ \bibinfo {pages} {1} (\bibinfo {year}
  {2012})}\BibitemShut {NoStop}%
\bibitem [{\citenamefont {Marchetti}\ \emph {et~al.}(2013)\citenamefont
  {Marchetti}, \citenamefont {Joanny}, \citenamefont {Ramaswamy}, \citenamefont
  {Liverpool}, \citenamefont {Prost}, \citenamefont {Rao},\ and\ \citenamefont
  {Simha}}]{Marchetti-2013}%
  \BibitemOpen
  \bibfield  {author} {\bibinfo {author} {\bibfnamefont {M.~C.}\ \bibnamefont
  {Marchetti}}, \bibinfo {author} {\bibfnamefont {J.-F.}\ \bibnamefont
  {Joanny}}, \bibinfo {author} {\bibfnamefont {S.}~\bibnamefont {Ramaswamy}},
  \bibinfo {author} {\bibfnamefont {T.~B.}\ \bibnamefont {Liverpool}}, \bibinfo
  {author} {\bibfnamefont {J.}~\bibnamefont {Prost}}, \bibinfo {author}
  {\bibfnamefont {M.}~\bibnamefont {Rao}}, \ and\ \bibinfo {author}
  {\bibfnamefont {R.~A.}\ \bibnamefont {Simha}},\ }\href@noop {} {\bibfield
  {journal} {\bibinfo  {journal} {Rev. Mod. Phys.}\ }\textbf {\bibinfo {volume}
  {85}},\ \bibinfo {pages} {1143} (\bibinfo {year} {2013})}\BibitemShut
  {NoStop}%
\bibitem [{\citenamefont {Cates}(2012)}]{Cates-2012}%
  \BibitemOpen
  \bibfield  {author} {\bibinfo {author} {\bibfnamefont {M.~E.}\ \bibnamefont
  {Cates}},\ }\href@noop {} {\bibfield  {journal} {\bibinfo  {journal} {Rep.
  Prog. Phys.}\ }\textbf {\bibinfo {volume} {75}},\ \bibinfo {pages} {042601}
  (\bibinfo {year} {2012})}\BibitemShut {NoStop}%
\bibitem [{\citenamefont {Howse}\ \emph {et~al.}(2007)\citenamefont {Howse},
  \citenamefont {Jones}, \citenamefont {Ryan}, \citenamefont {Gough},
  \citenamefont {Vafabakhsh},\ and\ \citenamefont {Golestanian}}]{Howse-2007}%
  \BibitemOpen
  \bibfield  {author} {\bibinfo {author} {\bibfnamefont {J.~R.}\ \bibnamefont
  {Howse}}, \bibinfo {author} {\bibfnamefont {R.~A.~L.}\ \bibnamefont {Jones}},
  \bibinfo {author} {\bibfnamefont {A.~J.}\ \bibnamefont {Ryan}}, \bibinfo
  {author} {\bibfnamefont {T.}~\bibnamefont {Gough}}, \bibinfo {author}
  {\bibfnamefont {R.}~\bibnamefont {Vafabakhsh}}, \ and\ \bibinfo {author}
  {\bibfnamefont {R.}~\bibnamefont {Golestanian}},\ }\href@noop {} {\bibfield
  {journal} {\bibinfo  {journal} {Phys. Rev. Lett.}\ }\textbf {\bibinfo
  {volume} {99}},\ \bibinfo {pages} {048102} (\bibinfo {year}
  {2007})}\BibitemShut {NoStop}%
\bibitem [{\citenamefont {Ebbens}\ and\ \citenamefont
  {Howse}(2010)}]{Ebbens-2010}%
  \BibitemOpen
  \bibfield  {author} {\bibinfo {author} {\bibfnamefont {S.~J.}\ \bibnamefont
  {Ebbens}}\ and\ \bibinfo {author} {\bibfnamefont {J.~R.}\ \bibnamefont
  {Howse}},\ }\href@noop {} {\bibfield  {journal} {\bibinfo  {journal} {Soft
  Matter}\ }\textbf {\bibinfo {volume} {6}},\ \bibinfo {pages} {726} (\bibinfo
  {year} {2010})}\BibitemShut {NoStop}%
\bibitem [{\citenamefont {Volpe}\ \emph {et~al.}(2011)\citenamefont {Volpe},
  \citenamefont {Buttinoni}, \citenamefont {Vogt}, \citenamefont {Kummerer},\
  and\ \citenamefont {Bechinger}}]{Volpe-2011}%
  \BibitemOpen
  \bibfield  {author} {\bibinfo {author} {\bibfnamefont {G.}~\bibnamefont
  {Volpe}}, \bibinfo {author} {\bibfnamefont {I.}~\bibnamefont {Buttinoni}},
  \bibinfo {author} {\bibfnamefont {D.}~\bibnamefont {Vogt}}, \bibinfo {author}
  {\bibfnamefont {H.-J.}\ \bibnamefont {Kummerer}}, \ and\ \bibinfo {author}
  {\bibfnamefont {C.}~\bibnamefont {Bechinger}},\ }\href@noop {} {\bibfield
  {journal} {\bibinfo  {journal} {Soft Matter}\ }\textbf {\bibinfo {volume}
  {7}},\ \bibinfo {pages} {8810} (\bibinfo {year} {2011})}\BibitemShut
  {NoStop}%
\bibitem [{\citenamefont {Thutupalli}\ \emph {et~al.}(2011)\citenamefont
  {Thutupalli}, \citenamefont {Seemann},\ and\ \citenamefont
  {Herminghaus}}]{Thutupalli-2011}%
  \BibitemOpen
  \bibfield  {author} {\bibinfo {author} {\bibfnamefont {S.}~\bibnamefont
  {Thutupalli}}, \bibinfo {author} {\bibfnamefont {R.}~\bibnamefont {Seemann}},
  \ and\ \bibinfo {author} {\bibfnamefont {S.}~\bibnamefont {Herminghaus}},\
  }\href@noop {} {\bibfield  {journal} {\bibinfo  {journal} {New J. Phys.}\
  }\textbf {\bibinfo {volume} {13}},\ \bibinfo {pages} {073021} (\bibinfo
  {year} {2011})}\BibitemShut {NoStop}%
\bibitem [{\citenamefont {Palacci}\ \emph {et~al.}(2013)\citenamefont
  {Palacci}, \citenamefont {Sacanna}, \citenamefont {Steinberg}, \citenamefont
  {Pine},\ and\ \citenamefont {Chaikin}}]{Palacci-2013}%
  \BibitemOpen
  \bibfield  {author} {\bibinfo {author} {\bibfnamefont {J.}~\bibnamefont
  {Palacci}}, \bibinfo {author} {\bibfnamefont {S.}~\bibnamefont {Sacanna}},
  \bibinfo {author} {\bibfnamefont {A.~P.}\ \bibnamefont {Steinberg}}, \bibinfo
  {author} {\bibfnamefont {D.~J.}\ \bibnamefont {Pine}}, \ and\ \bibinfo
  {author} {\bibfnamefont {P.~M.}\ \bibnamefont {Chaikin}},\ }\href@noop {}
  {\bibfield  {journal} {\bibinfo  {journal} {Science}\ }\textbf {\bibinfo
  {volume} {339}},\ \bibinfo {pages} {936} (\bibinfo {year}
  {2013})}\BibitemShut {NoStop}%
\bibitem [{\citenamefont {Narayan}\ \emph {et~al.}(2007)\citenamefont
  {Narayan}, \citenamefont {Ramaswamy},\ and\ \citenamefont
  {Menon}}]{Narayan-2007}%
  \BibitemOpen
  \bibfield  {author} {\bibinfo {author} {\bibfnamefont {V.}~\bibnamefont
  {Narayan}}, \bibinfo {author} {\bibfnamefont {S.}~\bibnamefont {Ramaswamy}},
  \ and\ \bibinfo {author} {\bibfnamefont {N.}~\bibnamefont {Menon}},\
  }\href@noop {} {\bibfield  {journal} {\bibinfo  {journal} {Science}\ }\textbf
  {\bibinfo {volume} {317}},\ \bibinfo {pages} {105} (\bibinfo {year}
  {2007})}\BibitemShut {NoStop}%
\bibitem [{\citenamefont {Deseigne}\ \emph {et~al.}(2010)\citenamefont
  {Deseigne}, \citenamefont {Dauchot},\ and\ \citenamefont
  {Chat\'e}}]{Deseigne-2010}%
  \BibitemOpen
  \bibfield  {author} {\bibinfo {author} {\bibfnamefont {J.}~\bibnamefont
  {Deseigne}}, \bibinfo {author} {\bibfnamefont {O.}~\bibnamefont {Dauchot}}, \
  and\ \bibinfo {author} {\bibfnamefont {H.}~\bibnamefont {Chat\'e}},\
  }\href@noop {} {\bibfield  {journal} {\bibinfo  {journal} {Phys. Rev. Lett.}\
  }\textbf {\bibinfo {volume} {105}},\ \bibinfo {pages} {098001} (\bibinfo
  {year} {2010})}\BibitemShut {NoStop}%
\bibitem [{\citenamefont {Zhang}\ \emph {et~al.}(2010)\citenamefont {Zhang},
  \citenamefont {Be{'}er}, \citenamefont {Florin},\ and\ \citenamefont
  {Swinney}}]{Zhang-2010}%
  \BibitemOpen
  \bibfield  {author} {\bibinfo {author} {\bibfnamefont {H.~P.}\ \bibnamefont
  {Zhang}}, \bibinfo {author} {\bibfnamefont {A.}~\bibnamefont {Be{'}er}},
  \bibinfo {author} {\bibfnamefont {E.-L.}\ \bibnamefont {Florin}}, \ and\
  \bibinfo {author} {\bibfnamefont {H.~L.}\ \bibnamefont {Swinney}},\
  }\href@noop {} {\bibfield  {journal} {\bibinfo  {journal} {Proc. Natl. Acad.
  Sci.}\ }\textbf {\bibinfo {volume} {107}},\ \bibinfo {pages} {13626}
  (\bibinfo {year} {2010})}\BibitemShut {NoStop}%
\bibitem [{\citenamefont {Wensink}\ and\ \citenamefont
  {L\"owen}(2012)}]{Wensink-2012}%
  \BibitemOpen
  \bibfield  {author} {\bibinfo {author} {\bibfnamefont {H.~H.}\ \bibnamefont
  {Wensink}}\ and\ \bibinfo {author} {\bibfnamefont {H.}~\bibnamefont
  {L\"owen}},\ }\href@noop {} {\bibfield  {journal} {\bibinfo  {journal} {J.
  Phys.: Condens. Matt.}\ }\textbf {\bibinfo {volume} {24}},\ \bibinfo {pages}
  {464130} (\bibinfo {year} {2012})}\BibitemShut {NoStop}%
\bibitem [{\citenamefont {Galajda}\ \emph {et~al.}(2007)\citenamefont
  {Galajda}, \citenamefont {Keymer}, \citenamefont {Chaikin},\ and\
  \citenamefont {Austin}}]{Galajda-2007}%
  \BibitemOpen
  \bibfield  {author} {\bibinfo {author} {\bibfnamefont {P.}~\bibnamefont
  {Galajda}}, \bibinfo {author} {\bibfnamefont {J.}~\bibnamefont {Keymer}},
  \bibinfo {author} {\bibfnamefont {P.}~\bibnamefont {Chaikin}}, \ and\
  \bibinfo {author} {\bibfnamefont {R.}~\bibnamefont {Austin}},\ }\href@noop {}
  {\bibfield  {journal} {\bibinfo  {journal} {J. Bacteriol.}\ }\textbf
  {\bibinfo {volume} {189}},\ \bibinfo {pages} {8704} (\bibinfo {year}
  {2007})}\BibitemShut {NoStop}%
\bibitem [{\citenamefont {Angelani}\ \emph {et~al.}(2009)\citenamefont
  {Angelani}, \citenamefont {Di~Leonardo},\ and\ \citenamefont
  {Ruocco}}]{Angelani-2009}%
  \BibitemOpen
  \bibfield  {author} {\bibinfo {author} {\bibfnamefont {L.}~\bibnamefont
  {Angelani}}, \bibinfo {author} {\bibfnamefont {R.}~\bibnamefont
  {Di~Leonardo}}, \ and\ \bibinfo {author} {\bibfnamefont {G.}~\bibnamefont
  {Ruocco}},\ }\href@noop {} {\bibfield  {journal} {\bibinfo  {journal} {Phys.
  Rev. Lett.}\ }\textbf {\bibinfo {volume} {102}},\ \bibinfo {pages} {048104}
  (\bibinfo {year} {2009})}\BibitemShut {NoStop}%
\bibitem [{\citenamefont {Lambert}\ \emph {et~al.}(2010)\citenamefont
  {Lambert}, \citenamefont {Liao},\ and\ \citenamefont
  {Austin}}]{Lambert-2010}%
  \BibitemOpen
  \bibfield  {author} {\bibinfo {author} {\bibfnamefont {G.}~\bibnamefont
  {Lambert}}, \bibinfo {author} {\bibfnamefont {D.}~\bibnamefont {Liao}}, \
  and\ \bibinfo {author} {\bibfnamefont {R.~H.}\ \bibnamefont {Austin}},\
  }\href@noop {} {\bibfield  {journal} {\bibinfo  {journal} {Phys. Rev. Lett.}\
  }\textbf {\bibinfo {volume} {104}},\ \bibinfo {pages} {168102} (\bibinfo
  {year} {2010})}\BibitemShut {NoStop}%
\bibitem [{\citenamefont {Pototsky}\ \emph {et~al.}(2013)\citenamefont
  {Pototsky}, \citenamefont {Hahn},\ and\ \citenamefont
  {Stark}}]{Potosky-2013}%
  \BibitemOpen
  \bibfield  {author} {\bibinfo {author} {\bibfnamefont {A.}~\bibnamefont
  {Pototsky}}, \bibinfo {author} {\bibfnamefont {A.~M.}\ \bibnamefont {Hahn}},
  \ and\ \bibinfo {author} {\bibfnamefont {H.}~\bibnamefont {Stark}},\
  }\href@noop {} {\bibfield  {journal} {\bibinfo  {journal} {Phys. Rev. E}\
  }\textbf {\bibinfo {volume} {87}},\ \bibinfo {pages} {042124} (\bibinfo
  {year} {2013})}\BibitemShut {NoStop}%
\bibitem [{\citenamefont {Schweitzer}\ and\ \citenamefont
  {Schimansky-Geier}(1994)}]{Schweitzer-1994}%
  \BibitemOpen
  \bibfield  {author} {\bibinfo {author} {\bibfnamefont {F.}~\bibnamefont
  {Schweitzer}}\ and\ \bibinfo {author} {\bibfnamefont {L.}~\bibnamefont
  {Schimansky-Geier}},\ }\href@noop {} {\bibfield  {journal} {\bibinfo
  {journal} {Physica A}\ }\textbf {\bibinfo {volume} {206}},\ \bibinfo {pages}
  {359} (\bibinfo {year} {1994})}\BibitemShut {NoStop}%
\bibitem [{\citenamefont {Cates}\ \emph {et~al.}(2010)\citenamefont {Cates},
  \citenamefont {Marenduzzo}, \citenamefont {Pagonabarraga},\ and\
  \citenamefont {Tailleur}}]{Cates-2010}%
  \BibitemOpen
  \bibfield  {author} {\bibinfo {author} {\bibfnamefont {M.~E.}\ \bibnamefont
  {Cates}}, \bibinfo {author} {\bibfnamefont {D.}~\bibnamefont {Marenduzzo}},
  \bibinfo {author} {\bibfnamefont {I.}~\bibnamefont {Pagonabarraga}}, \ and\
  \bibinfo {author} {\bibfnamefont {J.}~\bibnamefont {Tailleur}},\ }\href@noop
  {} {\bibfield  {journal} {\bibinfo  {journal} {Proc. Natl. Acad. Sci.}\
  }\textbf {\bibinfo {volume} {107}},\ \bibinfo {pages} {11715} (\bibinfo
  {year} {2010})}\BibitemShut {NoStop}%
\bibitem [{\citenamefont {Theurkauff}\ \emph {et~al.}(2012)\citenamefont
  {Theurkauff}, \citenamefont {Cottin-Bizonne}, \citenamefont {Palacci},
  \citenamefont {Ybert},\ and\ \citenamefont {Bocquet}}]{Theurkauff-2012}%
  \BibitemOpen
  \bibfield  {author} {\bibinfo {author} {\bibfnamefont {I.}~\bibnamefont
  {Theurkauff}}, \bibinfo {author} {\bibfnamefont {C.}~\bibnamefont
  {Cottin-Bizonne}}, \bibinfo {author} {\bibfnamefont {J.}~\bibnamefont
  {Palacci}}, \bibinfo {author} {\bibfnamefont {C.}~\bibnamefont {Ybert}}, \
  and\ \bibinfo {author} {\bibfnamefont {L.}~\bibnamefont {Bocquet}},\
  }\href@noop {} {\bibfield  {journal} {\bibinfo  {journal} {Phys. Rev. Lett.}\
  }\textbf {\bibinfo {volume} {108}},\ \bibinfo {pages} {268303} (\bibinfo
  {year} {2012})}\BibitemShut {NoStop}%
\bibitem [{\citenamefont {McCandlish}\ \emph {et~al.}(2012)\citenamefont
  {McCandlish}, \citenamefont {Baskaran},\ and\ \citenamefont
  {Hagan}}]{McCandlish-2012}%
  \BibitemOpen
  \bibfield  {author} {\bibinfo {author} {\bibfnamefont {S.~R.}\ \bibnamefont
  {McCandlish}}, \bibinfo {author} {\bibfnamefont {A.}~\bibnamefont
  {Baskaran}}, \ and\ \bibinfo {author} {\bibfnamefont {M.~F.}\ \bibnamefont
  {Hagan}},\ }\href@noop {} {\bibfield  {journal} {\bibinfo  {journal} {Soft
  Matter}\ }\textbf {\bibinfo {volume} {8}},\ \bibinfo {pages} {2527} (\bibinfo
  {year} {2012})}\BibitemShut {NoStop}%
\bibitem [{\citenamefont {Farrell}\ \emph {et~al.}(2012)\citenamefont
  {Farrell}, \citenamefont {Marchetti}, \citenamefont {Marenduzzo},\ and\
  \citenamefont {Tailleur}}]{Farrell-2012}%
  \BibitemOpen
  \bibfield  {author} {\bibinfo {author} {\bibfnamefont {F.~D.~C.}\
  \bibnamefont {Farrell}}, \bibinfo {author} {\bibfnamefont {M.~C.}\
  \bibnamefont {Marchetti}}, \bibinfo {author} {\bibfnamefont {D.}~\bibnamefont
  {Marenduzzo}}, \ and\ \bibinfo {author} {\bibfnamefont {J.}~\bibnamefont
  {Tailleur}},\ }\href@noop {} {\bibfield  {journal} {\bibinfo  {journal}
  {Phys. Rev. Lett.}\ }\textbf {\bibinfo {volume} {108}},\ \bibinfo {pages}
  {248101} (\bibinfo {year} {2012})}\BibitemShut {NoStop}%
\bibitem [{\citenamefont {Redner}\ \emph
  {et~al.}(2013{\natexlab{a}})\citenamefont {Redner}, \citenamefont
  {Baskaran},\ and\ \citenamefont {Hagan}}]{Redner-2013-PRE}%
  \BibitemOpen
  \bibfield  {author} {\bibinfo {author} {\bibfnamefont {G.~S.}\ \bibnamefont
  {Redner}}, \bibinfo {author} {\bibfnamefont {A.}~\bibnamefont {Baskaran}}, \
  and\ \bibinfo {author} {\bibfnamefont {M.~F.}\ \bibnamefont {Hagan}},\
  }\href@noop {} {\bibfield  {journal} {\bibinfo  {journal} {Phys. Rev. E}\
  }\textbf {\bibinfo {volume} {88}},\ \bibinfo {pages} {012305} (\bibinfo
  {year} {2013}{\natexlab{a}})}\BibitemShut {NoStop}%
\bibitem [{\citenamefont {Buttinoni}\ \emph {et~al.}(2013)\citenamefont
  {Buttinoni}, \citenamefont {Bialk\'e}, \citenamefont {K\"ummel},
  \citenamefont {L\"owen}, \citenamefont {Bechinger},\ and\ \citenamefont
  {Speck}}]{Buttinoni-2013}%
  \BibitemOpen
  \bibfield  {author} {\bibinfo {author} {\bibfnamefont {I.}~\bibnamefont
  {Buttinoni}}, \bibinfo {author} {\bibfnamefont {J.}~\bibnamefont {Bialk\'e}},
  \bibinfo {author} {\bibfnamefont {F.}~\bibnamefont {K\"ummel}}, \bibinfo
  {author} {\bibfnamefont {H.}~\bibnamefont {L\"owen}}, \bibinfo {author}
  {\bibfnamefont {C.}~\bibnamefont {Bechinger}}, \ and\ \bibinfo {author}
  {\bibfnamefont {T.}~\bibnamefont {Speck}},\ }\href@noop {} {\bibfield
  {journal} {\bibinfo  {journal} {Phys. Rev. Lett.}\ }\textbf {\bibinfo
  {volume} {110}},\ \bibinfo {pages} {238301} (\bibinfo {year}
  {2013})}\BibitemShut {NoStop}%
\bibitem [{\citenamefont {Tailleur}\ and\ \citenamefont
  {Cates}(2008)}]{Tailleur-2008}%
  \BibitemOpen
  \bibfield  {author} {\bibinfo {author} {\bibfnamefont {J.}~\bibnamefont
  {Tailleur}}\ and\ \bibinfo {author} {\bibfnamefont {M.~E.}\ \bibnamefont
  {Cates}},\ }\href@noop {} {\bibfield  {journal} {\bibinfo  {journal} {Phys.
  Rev. Lett.}\ }\textbf {\bibinfo {volume} {100}},\ \bibinfo {pages} {218103}
  (\bibinfo {year} {2008})}\BibitemShut {NoStop}%
\bibitem [{\citenamefont {Cates}\ and\ \citenamefont
  {Tailleur}(2013)}]{Cates-2013}%
  \BibitemOpen
  \bibfield  {author} {\bibinfo {author} {\bibfnamefont {M.~E.}\ \bibnamefont
  {Cates}}\ and\ \bibinfo {author} {\bibfnamefont {J.}~\bibnamefont
  {Tailleur}},\ }\href@noop {} {\bibfield  {journal} {\bibinfo  {journal}
  {EPL}\ }\textbf {\bibinfo {volume} {101}},\ \bibinfo {pages} {20010}
  (\bibinfo {year} {2013})}\BibitemShut {NoStop}%
\bibitem [{\citenamefont {Schnitzer}(1993)}]{Schnitzer-1993}%
  \BibitemOpen
  \bibfield  {author} {\bibinfo {author} {\bibfnamefont {M.~J.}\ \bibnamefont
  {Schnitzer}},\ }\href@noop {} {\bibfield  {journal} {\bibinfo  {journal}
  {Phys. Rev. E}\ }\textbf {\bibinfo {volume} {48}},\ \bibinfo {pages} {2553}
  (\bibinfo {year} {1993})}\BibitemShut {NoStop}%
\bibitem [{\citenamefont {Fily}\ and\ \citenamefont
  {Marchetti}(2012)}]{Fily-2012}%
  \BibitemOpen
  \bibfield  {author} {\bibinfo {author} {\bibfnamefont {Y.}~\bibnamefont
  {Fily}}\ and\ \bibinfo {author} {\bibfnamefont {M.~C.}\ \bibnamefont
  {Marchetti}},\ }\href@noop {} {\bibfield  {journal} {\bibinfo  {journal}
  {Phys. Rev. Lett.}\ }\textbf {\bibinfo {volume} {108}},\ \bibinfo {pages}
  {235702} (\bibinfo {year} {2012})}\BibitemShut {NoStop}%
\bibitem [{\citenamefont {Redner}\ \emph
  {et~al.}(2013{\natexlab{b}})\citenamefont {Redner}, \citenamefont {Hagan},\
  and\ \citenamefont {Baskaran}}]{Redner-2013}%
  \BibitemOpen
  \bibfield  {author} {\bibinfo {author} {\bibfnamefont {G.~S.}\ \bibnamefont
  {Redner}}, \bibinfo {author} {\bibfnamefont {M.~F.}\ \bibnamefont {Hagan}}, \
  and\ \bibinfo {author} {\bibfnamefont {A.}~\bibnamefont {Baskaran}},\
  }\href@noop {} {\bibfield  {journal} {\bibinfo  {journal} {Phys. Rev. Lett.}\
  }\textbf {\bibinfo {volume} {110}},\ \bibinfo {pages} {055701} (\bibinfo
  {year} {2013}{\natexlab{b}})}\BibitemShut {NoStop}%
\bibitem [{SI()}]{SI}%
  \BibitemOpen
  \href@noop {} {}\bibinfo {note} {See Supplemental Material at [URL] for
  movies, additional figures and simulation details.}\BibitemShut {Stop}%
\bibitem [{\citenamefont {Fielding}(2012)}]{Fielding-2012}%
  \BibitemOpen
  \bibfield  {author} {\bibinfo {author} {\bibfnamefont {S.~M.}\ \bibnamefont
  {Fielding}},\ }\href@noop {} {\bibfield  {journal} {\bibinfo  {journal}
  {arXiv:1210.5464}\ } (\bibinfo {year} {2012})}\BibitemShut {NoStop}%
\bibitem [{\citenamefont {Bialk\'e}\ \emph {et~al.}(2013)\citenamefont
  {Bialk\'e}, \citenamefont {L\"owen},\ and\ \citenamefont
  {Speck}}]{Bialke-2013}%
  \BibitemOpen
  \bibfield  {author} {\bibinfo {author} {\bibfnamefont {J.}~\bibnamefont
  {Bialk\'e}}, \bibinfo {author} {\bibfnamefont {H.}~\bibnamefont {L\"owen}}, \
  and\ \bibinfo {author} {\bibfnamefont {T.}~\bibnamefont {Speck}},\
  }\href@noop {} {\bibfield  {journal} {\bibinfo  {journal} {EPL}\ }\textbf
  {\bibinfo {volume} {103}},\ \bibinfo {pages} {30008} (\bibinfo {year}
  {2013})}\BibitemShut {NoStop}%
\bibitem [{\citenamefont {Plimpton}(1995)}]{Plimpton-1995}%
  \BibitemOpen
  \bibfield  {author} {\bibinfo {author} {\bibfnamefont {S.}~\bibnamefont
  {Plimpton}},\ }\href@noop {} {\bibfield  {journal} {\bibinfo  {journal} {J.
  Comput. Phys.}\ }\textbf {\bibinfo {volume} {117}},\ \bibinfo {pages} {1}
  (\bibinfo {year} {1995})}\BibitemShut {NoStop}%
\bibitem [{\citenamefont {Velasco}\ and\ \citenamefont
  {Toxv{\ae}rd}(1996)}]{Velasco-1996}%
  \BibitemOpen
  \bibfield  {author} {\bibinfo {author} {\bibfnamefont {E.}~\bibnamefont
  {Velasco}}\ and\ \bibinfo {author} {\bibfnamefont {S.}~\bibnamefont
  {Toxv{\ae}rd}},\ }\href@noop {} {\bibfield  {journal} {\bibinfo  {journal}
  {Phys. Rev. E}\ }\textbf {\bibinfo {volume} {54}},\ \bibinfo {pages} {605}
  (\bibinfo {year} {1996})}\BibitemShut {NoStop}%
\bibitem [{\citenamefont {Novik}\ and\ \citenamefont
  {Coveney}(2000)}]{Novik-2000}%
  \BibitemOpen
  \bibfield  {author} {\bibinfo {author} {\bibfnamefont {K.~E.}\ \bibnamefont
  {Novik}}\ and\ \bibinfo {author} {\bibfnamefont {P.~V.}\ \bibnamefont
  {Coveney}},\ }\href@noop {} {\bibfield  {journal} {\bibinfo  {journal} {Phys.
  Rev. E}\ }\textbf {\bibinfo {volume} {61}},\ \bibinfo {pages} {435} (\bibinfo
  {year} {2000})}\BibitemShut {NoStop}%
\bibitem [{\citenamefont {Chaikin}\ and\ \citenamefont
  {Lubensky}(1995)}]{Chaikin}%
  \BibitemOpen
  \bibfield  {author} {\bibinfo {author} {\bibfnamefont {P.}~\bibnamefont
  {Chaikin}}\ and\ \bibinfo {author} {\bibfnamefont {T.~C.}\ \bibnamefont
  {Lubensky}},\ }\href@noop {} {\emph {\bibinfo {title} {Principles of
  condensed matter physics}}}\ (\bibinfo  {publisher} {Cambridge University
  Press},\ \bibinfo {year} {1995})\BibitemShut {NoStop}%
\bibitem [{\citenamefont {Bray}(2002)}]{Bray-2002}%
  \BibitemOpen
  \bibfield  {author} {\bibinfo {author} {\bibfnamefont {A.~J.}\ \bibnamefont
  {Bray}},\ }\href@noop {} {\bibfield  {journal} {\bibinfo  {journal} {Adv.
  Phys.}\ }\textbf {\bibinfo {volume} {51}},\ \bibinfo {pages} {481} (\bibinfo
  {year} {2002})}\BibitemShut {NoStop}%
\bibitem [{\citenamefont {Redner}()}]{Redner-email}%
  \BibitemOpen
  \bibfield  {author} {\bibinfo {author} {\bibfnamefont {G.}~\bibnamefont
  {Redner}},\ }\href@noop {} {}\bibinfo {note} {Private
  {c}ommunication}\BibitemShut {NoStop}%
\bibitem [{\citenamefont {Bray}\ and\ \citenamefont
  {Emmott}(1995)}]{Bray-1995}%
  \BibitemOpen
  \bibfield  {author} {\bibinfo {author} {\bibfnamefont {A.~J.}\ \bibnamefont
  {Bray}}\ and\ \bibinfo {author} {\bibfnamefont {C.~L.}\ \bibnamefont
  {Emmott}},\ }\href@noop {} {\bibfield  {journal} {\bibinfo  {journal} {Phys.
  Rev. B}\ }\textbf {\bibinfo {volume} {52}},\ \bibinfo {pages} {R685}
  (\bibinfo {year} {1995})}\BibitemShut {NoStop}%
\end{thebibliography}%

\end{document}